\documentclass[aps,pra,superscriptaddress]{revtex4}
\usepackage{graphicx}
\usepackage{enumerate}

\newcommand{\ket}[1]{| \, #1 \rangle}
\newcommand{\bra}[1]{ \langle #1 \,  |}
\newcommand{\proj}[1]{\ket{#1}\bra{#1}}

\newcommand{\abs}[1]{ | \, #1 \,  |}

\newcommand{\braket}[2]{\langle #1 \,|\, #2 \rangle}

\newcommand{\beq}{\begin{equation}}
\newcommand{\eeq}{\end{equation}}

\begin{document}
\title{Quantum and classical correlations between players in game theory}

\author{Junichi Shimamura}
\email{shimamura_junichi@soken.ac.jp}
\author{\c{S}ahin Kaya \"{O}zdemir}
\affiliation{CREST Research Team for Interacting Carrier
Electronics, The Graduate University for Advanced Studies
(SOKENDAI), Hayama, Kanagawa 240-0193, Japan }
\author{Fumiaki Morikoshi}
\affiliation{NTT Basic Research Laboratories, NTT Corporation, 3-1
Morinosato Wakamiya, Atsugi, Kanagawa 243-0198, Japan}
\author{Nobuyuki Imoto}
\affiliation{CREST Research Team for Interacting Carrier
Electronics, The Graduate University for Advanced Studies
(SOKENDAI), Hayama, Kanagawa 240-0193, Japan } \affiliation{NTT
Basic Research Laboratories, NTT Corporation, 3-1 Morinosato
Wakamiya, Atsugi, Kanagawa 243-0198, Japan}

\date{\today}

\begin{abstract}
Effects of quantum and classical correlations on game theory are
studied to clarify the new aspects brought into game theory by
the quantum mechanical toolbox. In this study, we compare 
quantum correlation represented by a maximally entangled state
and classical correlation that is generated through phase damping
processes on the maximally entangled state. Thus, this also sheds
light on the behavior of games under the influence of noisy sources.
It is observed that the quantum correlation can always resolve the
dilemmas in non-zero sum games and attain the maximum sum of both
 players' payoffs, while the classical correlation cannot necessarily
resolve the dilemmas.
\end{abstract}

\maketitle

\section{Introduction}

Recently, there has been a growing interest in game theory within the
quantum information community \cite{Eisert1,Meyer}. The concepts of
quantum 
mechanics have been introduced to study quantum versions of classical game theory. This new version, which is referred to as
{\it Quantum Game Theory}, can exploit quantum superpositions,
entanglement, and quantum operations that are absent from classical game
theory. So far, researches on the quantum versions of game
theory have focused mainly on observing the dynamics and
outcomes of games in the quantum domain. One of the motivations of
introducing quantum mechanics into game theory is to find
a solution resolving the dilemmas in the games, which cannot be achieved in
classical game theory. Another one is that the study of
quantum versions of game theory might be useful to
examine many complicated and open problems in quantum
information science.

Most of the studies that have been conducted so far are devoted to
resolving the dilemmas inherit in classical games with the help
of quantum mechanics. In fact, they have been resolved by using
entanglement. In the course of the research, however, we come up with the question:
{\it Is entanglement really essential to resolve the dilemmas in
the games ?} In other words, {\it can classical correlations 
replace quantum ones (entanglement) to obtain
similar results ?} If entanglement turns out to be inessential, it is
natural to ask {\it what the difference in payoffs depending on 
the types of correlations is}.

Our purpose in this study is to make a comparative analysis of the
effects of shared quantum and classical correlations on the
strategies of players and on the dynamics of games.
In particular, we would like to focus on whether classical
correlations can produce the same results as that of quantum
correlations. Even if they cannot, it is worth investigating to what extent classical 
correlations can enhance original classical games.
In this study, we consider the
following games: Prisoner's Dilemma (PD), Chicken Game (CG), Stag
Hunt (SH), Battle of Sexes (BoS), Matching Penny (MP), Samaritan's
Dilemma (SD), and the Monty Hall game (MH). All of these games have
different types of dilemmas and they fall into different classes
when they are classified according to their classical payoff
matrices. All but MH, which is a sequential game, are 2$\times$2
strategic games (see Ref. \cite{Rasmusen} for details of these games).

\section{Classical game theory}

\begin{figure}[tbp]
\begin{center}
 \includegraphics[scale=0.65]{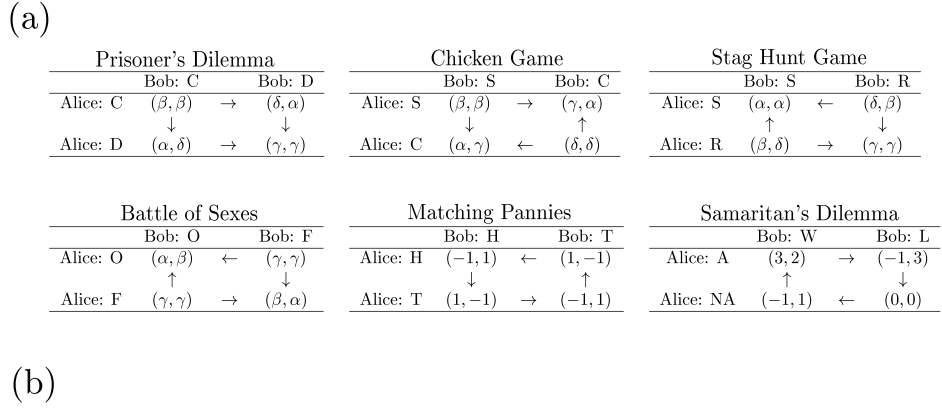}
\end{center}
\caption{(a) Payoff matrices for games in this study where $\alpha > \beta > \gamma > \delta$ and $2 \beta>\alpha + \delta$ for symmetric games. Additional conditions $\alpha + \delta > \beta + \gamma$ for PD, $\alpha + \delta= \beta + \gamma$ for CG, and $\alpha + \delta > \beta + \gamma$ for SH are imposed. Arrows show the best response of a player to a fixed choice of the other player. Any strategy pointed by two arrows is an NE. (b) Schematic configuration of the quantization scheme for $2\times 2$ strategic games. The operations inside the dotted boxes are performed by the referee.}
\label{fig:fig1}
\end{figure}

In this section we will briefly describe the rules of classical game
theory and introduce several games studied in this paper.
A strategic game can be denoted by $\Gamma=[N, (S_{i})_{i \in N}, (u_{i})_{i \in N}]$ where $N$ is the set of players,
$S_{i}$ is the set of strategies for the $i$-th player, and $u_{i}$ is
the payoff function for the $i$-th player, which is a map from the set of all possible strategy combinations
into real numbers \cite{myerson}. Then the payoff for the $i$-th player
can be denoted as $u_{i}(s)$ where $s$ is a combination of the strategies implemented by all players.
In classical game theory, any strategic game is fully described by
its payoff matrix. For 2$\times$2 strategic games, the properties of the
games are
determined by each player's four possible payoffs. The payoff
matrices of the strategic games we discuss are shown in Fig.\ref{fig:fig1}(a).
2$\times$2 games can be classified into 78 types, provided that all the four
possible payoffs for each player are different \cite{Rapop}. 
We will deal with only games that have dilemmas, while most of
these 78 different types do not have dilemmas.

According to the above classification into 78
types, there are only four symmetric games with a dilemma which
arises from the incentive not to cooperate. These four games are
PD, CG, SH, and Deadlock game \cite{poundstone}. 
In the last one, players do not face the problem of choosing the strategy because players can decide on a unique Nash equilibria (NE) without
hesitation in the sense that they do not have any incentive to deviate from
the NE. 
On the other hand, for PD and SH, there exist incentives to deviate from
an NE from mathematical or psychological motivation. For CG, since there exist
two equivalent NE's, players cannot decide on which NE to choose. 

Next, we will briefly mention the nature of the dilemma in each game.
As it can be seen from the tables, PD has one NE where both players get the payoff $\gamma$ when they
both choose to play ``defect'' (D). On the other hand, when both
play ``cooperate'' (C), they get the payoff $\beta$ corresponding
to Pareto optimal. Here the dilemma occurs because Pareto
optimal and the NE of the game do not coincide.

The type of the dilemma in CG is different
from that in PD. In CG, the dilemma occurs because there exist 
two NE's with the payoffs $(\alpha, \gamma)$ and $(\gamma,\alpha)$ at the strategy sets (C,S) and (S,C) where S and C stand for
``swerve'' and ``continue'', respectively. Therefore, the players
without communicating with each other cannot decide on which NE to choose.

The type of the dilemma in SH is very different from those in the above two games.
SH has two NE's at which both players go for a ``stag'' (S) or go for a
``rabbit'' (R). Obviously, going for a ``stag'' is the best strategy for
both players.
For PD and CG, games are discussed on the assumption that players are
rational, on which there exists no dilemma in SH.
However, in this game, the dilemma can be found in the following sense:
If one goes for a stag but the other player goes for a rabbit, then the
former get the worst possible payoff $\delta$, while the latter still
get the second highest payoff $\beta$. In other words, the dilemma arises from
the fear that the other player might not be rational. This type of the
dilemma can often be seen in society \cite{poundstone}.

Having shown the dilemmas in symmetric games, we will explain asymmetric
games below.
BoS has a dilemma similar to that in CG. The
NE's are at (O,O) and (F,F) where O is the choice of going to
``opera'' and F is the choice of going to a ``football match''. The
difference between BoS and CG is that the former has an asymmetric
payoff matrix, whereas the latter has a symmetric one. BoS and the three
games mentioned above have at least one NE.

MP is different from the above four games,
not only because it is a zero sum game, but also because it is a {\it
discoordination game} \cite{Rasmusen}, i.e., there is no NE in the
original classical game. Since the interests of players conflict,
they will never try to coordinate their strategies of playing ``head''
(H) or ``tail'' (T).
This game is also classified as an asymmetric game.

SD is also a discoordination and asymmetric
game but not a zero sum game. It is seen from the table that there
is no NE in the pure strategies. When the players adopt classical
mixed strategies, a unique NE is realized
 when Alice chooses ``aid'' (A) with probability 0.5 and Bob chooses ``work'' (W) with
probability 0.25. Although the classical mixed strategy gives the
unique NE, Alice's payoff is negative. This is the dilemma in the game,
which corresponds to the fact that Alice, who likes to help the person in
need (Bob) voluntarily, is exploited by the selfish behavior of Bob and she
cannot stop this exploitation. In the classical mixed strategies,
the payoff of Alice and Bob are -0.25 and 1.5, respectively, where
there still remains the dilemma for Alice because her payoff is
negative, implying her loss of resources.

MH is classified as a sequential game \cite{monty}. The scenario is as
follows: The participant (Bob) is given the opportunity to select
one out of three closed doors, behind one of which there is a prize.
There is no prize behind the other two doors. Once he makes his
choice, Monty Hall (Alice) will open one of the remaining
doors, revealing that it does not contain the prize. She then asks
him if he would like to switch his choice to the other unopened
door, or stay with his original choice. Can he increase the probability
of wining the prize by changing his choice ?  MH attracted much interest because the answer is
counterintuitive. Classically, the best strategy for Bob is to
switch his choice, which doubles the chance of wining the prize.

\section{Scheme for the quantum version of $2\times2$ games}

\subsection{Quantum operations (QO) and quantum correlation (QC)}
We will introduce the scheme for a quantum version of these games.
A quantum version of a 2$\times$2 classical game with a
maximally entangled state shared between the players is played
as follows \cite{Eisert1}[see Fig.\ref{fig:fig1} (b)]: (a) A referee prepares a maximally entangled state by
applying an entangling operator $\hat{J}$ on a product state
$\ket{f}\ket{g}$ where $\hat{J}$ is defined as
$\hat{J}\ket{f}\ket{g}=\frac{1}{\sqrt{2}} \left[ \ket{f}\ket{g} + i (-1)^{(f+g)}\ket{(1-f)}\ket{(1-g)} \right]$ with $f,g = 0,1$. The output
of this entangler $\rho_{in}=\hat{J}\ket{fg}\bra{fg}\hat{J}^{\dagger}$ is delivered
to the players. (b) The players apply their actions, which are
SU(2) operators, locally on their qubits, and return the resultant
state $\rho_{out}= ( \hat{U}_{A} \otimes \hat{U}_{B} ) \rho_{in} (\hat{U}^{\dagger}_{A} \otimes \hat{U}^{\dagger}_{B} )$ to the
referee. Alice's and Bob's actions are restricted to the
two-parameter SU(2) operators, $\hat{U}_{A}(\theta_{A},\phi_{A})$
and $\hat{U}_{B}(\theta_{B},\phi_{B})$, given as 
\beq\label{Eq1}
\hat{U}(\theta,\phi)=\left( \begin{array}{cc}
       e^{i \phi}\cos\frac{\theta}{2} & \sin\frac{\theta}{2} \\
        \sin\frac{\theta}{2}       &  e^{-i \phi}\cos\frac{\theta}{2}\end{array} \right),
\eeq 
with $0 \leq \phi \leq \pi/2$ and $0 \leq \theta \leq \pi$.
In particular, the identity operator ($\hat{\sigma}_{0}$) and the
bit flip operator ($i \hat{\sigma}_{y}$) correspond to the two classical
pure strategies.
(c) The referee, upon receiving this state, applies
$\hat{J}^{\dagger}$ and then makes a measurement $\{\Pi_{n}=\ket{jl}\bra{jl} \}_{ \{ {j,l}={0,1} \} }$ with $n=2j+l$.
According to the measurement outcome $n$, the referee assigns each
player the payoff chosen from the payoff matrix of the original classical
game. Then the average payoff of the players can be written as 
$\$_{A} = \sum_{n}a_{n}\rm{Tr}(\Pi_{n}\it{\hat{J}}^{\dagger}\hat{\rho}_{\rm{out}}\hat{J}),
\$_{B}=\sum_{n} b_{n}\rm{Tr} (\Pi_{n}\it{\hat{J}}^{\dagger}\hat{\rho}_{\rm{out}}\hat{J})$ where $a_{n}$ and $b_{n}$ are the payoffs chosen from the classical
payoff matrix when the measurement result is $n$.

When an input state is $\ket{f}\ket{g}=\ket{0}\ket{0}$, the probabilities of obtaining $n=2j+l$, $P_{jl}$'s, are given by 
\begin{eqnarray}
P_{00} & = & \abs{\cos(\theta_{A}/2) \cos(\theta_{B}/2) \cos(\phi_{A} + \phi_{B})}^2, \nonumber \\
P_{01} & = & \abs{x\sin\phi_{B} - y \cos\phi_{A}}^2, \nonumber \\
P_{10} & = & \abs{x\cos\phi_{B} - y \sin\phi_{A}}^2,  \nonumber \\
P_{11} & = & \abs{\sin(\theta_{A}/2)\sin(\theta_{B}/2) + \cos(\theta_{A}/2)\cos(\theta_{B}/2)\sin(\phi_{A} + \phi_{B})}^2,
\end{eqnarray}
where $x=\sin(\theta_{A}/2)\cos(\theta_{B}/2)$ and $y=\cos(\theta_{A}/2)\sin(\theta_{B}/2)$.

\subsection{Quantum operations (QO) and classical correlation (CC)}
Here we assume that the players are allowed to use quantum
operations as shown in Eq.(\ref{Eq1}), however, the shared state
is a classically correlated state of the form
$\rho'=\frac{1}{2}[\ket{00}\bra{00} + \ket{11}\bra{11}]$. This can
occur, for example, when the maximally entangled state generated
by the referee undergoes a damping process where the off-diagonal
elements of the density operator disappear.
 Thus the game with a shared classical correlation
proceeds as in the preceding subsection with $\rho_{in}$ replaced by
a classically correlated state $\rho'$.

When $\rho'=\frac{1}{2}[\ket{00}\bra{00} + \ket{11}\bra{11}]$, the
probabilities of obtaining $n=2j+l$, $P_{jl}$'s, are given by
\begin{eqnarray}
P_{00} & = & \frac{1}{4}[1 + \cos\theta_{A}\cos\theta_{B} - \sin\theta_{A}\sin\theta_{B}\sin(\phi_{A}+\phi_{B})], \nonumber \\
P_{01} & = & \frac{1}{4}[1 - \cos\theta_{A}\cos\theta_{B} + \sin\theta_{A}\sin\theta_{B}\sin(\phi_{A}-\phi_{B})], \nonumber \\
P_{10} & = & \frac{1}{4}[1 - \cos\theta_{A}\cos\theta_{B} - \sin\theta_{A}\sin\theta_{B}\sin(\phi_{A}-\phi_{B})],  \nonumber \\
P_{11} & = & \frac{1}{4}[1 + \cos\theta_{A}\cos\theta_{B} + \sin\theta_{A}\sin\theta_{B}\sin(\phi_{A}+\phi_{B})].
\end{eqnarray}

\section{Effects of shared quantum and classical correlations on the dynamics of the strategic games}

In this section, we make a comparative analysis of the effects of
shared quantum and classical correlations on the various types of
games: (i) symmetric, non-zero sum game, (ii) asymmetric, non-zero
sum game, (iii) asymmetric, zero sum, discoordination game, and
(iv) asymmetric, non-zero sum, discoordination game.

\subsection{Symmetric, Non-zero sum game (PD, CG, SH)}
In symmetric games, payoff functions are symmetric, which means Bob's
payoff can be calculated from the payoff function of Alice by
interchanging Alice's and Bob's strategies. In PD, Alice's payoff
is given by
\begin{eqnarray}
\$_{A} &=& \alpha P_{10} + \beta P_{00} + \gamma P_{11} + \delta P_{01} \nonumber \\
       &=& (\alpha - \delta) P_{10} + (\beta - \delta) P_{00} + (\gamma - \delta) P_{11} + \delta.
\label{pf:eqn}
\end{eqnarray}
For CG, her payoff can be obtained as in Eq. (\ref{pf:eqn}) by
interchanging $\gamma$ and  $\delta$. For SH, her
payoff is given as in Eq. (\ref{pf:eqn}) by interchanging $\alpha$ and
$\beta$.

\textbf{\it{QO with QC:}}
When PD is played with a shared quantum correlation, it has already been
known that the strategy $(\hat{U}_{A}=i \hat{\sigma}_{z},
\hat{U}_{B}=i \hat{\sigma}_{z})$ gives the payoff
$(\$_{A},\$_{B})=(3,3)$ which coincides with Pareto optimal \cite{Eisert1}. Thus,
the dilemma in this game is resolved. In CG, the coefficient of
$P_{10}$ in Alice's payoff is greater than the others and thus the
same argument as in PD can be applied. The strategy
$(\hat{U}_{A}= i \hat{\sigma}_{z}, \hat{U}_{B}=i \hat{\sigma}_{z})$
gives a unique NE and can resolve the dilemma. 
For SH, there exist two equivalent NE's with payoffs
$(\$_{A},\$_{B})=(\alpha,\alpha)$ where $(\hat{U}_{A}=\hat{\sigma}_{0},\hat{U}_{B}= \hat{\sigma}_{0})$ or $(\hat{U}_{A}=i \hat{\sigma}_{z},\hat{U}_{B}=i \hat{\sigma}_{z})$.
The dilemma in SH comes from the fear that if one chases a stag and
the other chases a rabbit, the former cannot get anything. 
The former strategy corresponds to (S,S), however this cannot resolve the fear,
i.e., if one player choose $i \hat{\sigma}_{y}$ (R), the other get the
worst possible payoff, $\delta$. On the other hand, the latter strategy
realizes an NE with the best possible payoffs for both
players. Furthermore it guarantees a player choosing the strategy $i
\hat{\sigma}_{z}$ at least the second lowest payoff $\gamma$,  whatever strategy the other player chooses. It follows
that the strategy $(\hat{U}_{A}=i \hat{\sigma}_{z},\hat{U}_{B}=i
\hat{\sigma}_{z})$ 
is the best for the players and it resolves the dilemma in this game.
Consequently, it turns out that quantum correlations and
quantum operations can resolve the dilemma of ``Symmetric,
Non-zero sum games'' with the same strategy,
$\hat{U}_{A}=\hat{U}_{B}=i\hat{\sigma}_{z}$.

\textbf{\it{QO with CC:}} 
We analyze the effects of the classical
correlations on the dynamics of the games. Let us begin with PD, in
which Alice's payoff is given by
\begin{eqnarray}
\$_{A} & = & \frac{1}{4}[(\alpha + \beta + \gamma + \delta) + (- \alpha + \beta + \gamma - \delta)\cos\theta_{A}\cos\theta_{B} \nonumber \\
       &  &  + \sin\theta_{A}\sin\theta_{B}\{ (- \alpha - \beta + \gamma + \delta)\sin\phi_{A}\cos\phi_{B} \nonumber \\
       &  &  + (\alpha - \beta + \gamma - \delta)\cos\phi_{A}\sin\phi_{B}
       \}].
\end{eqnarray}
Here, we impose the condition $\alpha + \delta > \beta + \gamma$, which
is common to the conventional payoff matrices of PD with specific values. The analysis reveals the existence of three NE's, two of
which give the same amount of payoffs while the third NE gives
much smaller payoff to the players. The two NE's with equal
payoffs emerge at $(\theta_{A}=0,\forall
\phi_{A},\hat{U}_{B}=i\hat{\sigma}_{y})$ and
$(\hat{U}_{A}=i\hat{\sigma}_{y},\theta_{B}=0,\forall \phi_{B})$. The other NE
with a smaller payoff appears at $(\theta_{A}=\pi /2,\forall
\phi_{A},\theta_{B}=\pi /2,\forall \phi_{B})$.
In CG, we impose the condition $\alpha + \delta = \beta + \gamma$, which is common to the conventional payoff matrices of CG with specific values. The situation remains the same as PD except for the strategy that realizes the third NE. Thus, the dilemma cannot be resolved in this case as well.
In SH, there exist three NE's which are the same as those in PD;
therefore the dilemma cannot be resolved due to the existence of
two NE's giving the same payoffs.
Consequently, it follows that classical correlations cannot
realize any NE achieved by quantum correlations and thus cannot
resolve the dilemma.  In the case of classical correlations, the
main obstacle is the existence of two NE's with the same payoff.
Therefore, the players cannot decide on which NE to choose.

\subsection{Asymmetric, Non-zero sum game (BoS)}

In this game, Alice's payoff is given as
\begin{eqnarray}
\$_{A} &=& \alpha P_{00} + \beta P_{11} + \gamma (P_{01} +
P_{10})=(\alpha - \gamma) P_{00} + (\beta - \gamma) P_{11} +
\gamma ,
\end{eqnarray}
from which Bob's payoff can be obtained by interchanging $\alpha$
and $\beta$.

\textbf{\it{QO with QC:}}  
In BoS, classical mixed strategies where Alice and Bob chooses
``O''  with probabilities $(\frac{\alpha - \gamma}{\alpha + \beta
-2 \gamma},\frac{\beta - \gamma}{\alpha + \beta -2 \gamma})$ or
$(\frac{\beta - \gamma}{\alpha + \beta -2 \gamma},\frac{\alpha -
\gamma}{\alpha + \beta -2 \gamma})$ give NE's with payoff
$(\$_{A},\$_{B})=(\frac{\alpha \beta - \gamma}{\alpha + \beta -2\gamma},\frac{\alpha \beta - \gamma}{\alpha + \beta - 2 \gamma})$.
On the other hand, using QO with QC,
the players have infinite number of strategies
resulting in infinite number of NE's with the same payoff
$(\$_{A},\$_{B})=(\beta - \gamma,\alpha - \gamma )$ where
$(\theta_{A}=\theta_{B}, \phi_{A}=\frac{\pi}{2} - \phi_{B})$ and
$(\hat{U}_{A}=i \hat{\sigma}_{y}, \hat{U}_{B}=i
\hat{\sigma}_{y})$. Since the payoffs are all equal in the NE's,
the players cannot decide on which one to choose. But the concept of the
focal point effect \cite{Rasmusen} helps the players resolve the dilemma. The
strategy $(\hat{U}_{A}=i \hat{\sigma}_{y}, \hat{U}_{B}=i \hat{\sigma}_{y})$ is
distinguished from other strategies because the players do not
have to be concerned with the choice of the phase factor. Bob's
payoff is higher than Alice's at all of these NE's. This is an NE
with unbalanced payoffs for Alice and Bob. While this unbalanced
situation obtained when $\ket{f}\ket{g}=\ket{0}\ket{0}$ is
favorable for Bob, one can easily show that if they start with
$\ket{f}\ket{g}=\ket{1}\ket{1}$, this unbalanced situation will be
favorable for Alice, that is, her payoff will be higher than Bob's.
This simple example shows the dependence of players' payoffs on the
input state.

\textbf{\it{QO with CC:}} 
In this case, we find two NE's given by $(\theta_{A}=0,\forall
\phi_{A}, \theta_{B}=0,\forall \phi_{B})$ and $(\hat{U}_{A}=i
\hat{\sigma}_{y},\hat{U}_{B}=i \hat{\sigma}_{y})$ with the same payoffs
$(\$_{A},\$_{B})=((\alpha + \beta)/2,(\alpha + \beta)/2)$. The
players, however, cannot decide on which NE's to choose and thus the
dilemma cannot be resolved. As a result, one can say that in
this type of games, quantum correlations can achieve the unique NE
with the help of the concept of the focal point, while classical
correlations cannot.

\subsection{Asymmetric, Zero sum, Discoordination game (MP)}
For MP, Alice's payoff is given by
\begin{eqnarray}
\$_{A} &=& - P_{00} - P_{11} + P_{01} + P_{10} = 1-2(P_{00}
+P_{11}). 
\end{eqnarray}
Since this is a zero sum game, Bob's payoff is obtained as $\$_{B}=-\$_{A}$
. Note that, in zero sum games, it is not the purpose of the players to
find an equilibrium point on which they can compromise but rather to win the game and
beat the other player. A straightforward calculation reveals
that the players cannot find an equilibrium point when they use
quantum operations and quantum correlation. On the other hand,
when they play the game with quantum operations and classical
correlation, there emerge NE's with equal payoffs
$(\$_{A},\$_{B})=(0,0)$ when the players apply the operators
$(\theta_{A}=\frac{\pi}{2},\forall \phi_{A},
\theta_{B}=\frac{\pi}{2},\forall \phi_{B})$. In this case, there
is no winner and loser in the game. The payoffs they receive are
equal to those they receive when they play the game with classical
mixed strategies without any shared correlation.

\subsection{Asymmetric, Non-zero sum, Discoordination game (SD)}
Alice's and Bob's payoffs in SD are given by,
\begin{eqnarray}
\$_{A} &=& 3 P_{00} - P_{01} - P_{10}, \nonumber \\
\$_{B} &=& 2 P_{00} + 3 P_{01} + P_{10}.
\end{eqnarray}

\textbf{\it{QO with QC:}}  
Our analysis reveals that there is the unique NE with the payoff
given as $(3,2)$ where $(\hat{U}_{A}=i
\hat{\sigma}_{z},\hat{U}_{B}= i \hat{\sigma}_{z})$. Introducing
quantum operations and quantum correlation results in the
emergence of the unique NE which cannot be seen in classical
strategies. Both players receive higher payoffs than those
obtained when a classical mixed strategy is applied. Note that, in
classical mixed strategies, payoffs of the players are $(-0.2,1.5)$,
which is much smaller than that obtained with QO and QC.

\textbf{\it{QO with CC:}}  
We found that there appears a unique NE with the payoff
$(0.25,1.5)$ when the players apply $(\forall \theta_{A},
\phi_{A}=0, \hat{U}_{B}=\frac{1}{\sqrt{2}}(\hat{\sigma}_{0} + i
\hat{\sigma}_{y}))$. The payoff for Alice becomes greater than
that of the classical mixed strategy while Bob's remains unchanged.

Since Alice's payoff $\$_{A}$ becomes positive in both cases with quantum and classical correlations, the dilemma in this game is successfully resolved. 
However, if we impose the further condition, $\$_{A} > \$_{B}$, then it
can be achieved only with the quantum correlation. Although the case of
the classical correlation gives a unique NE, it cannot provide a
solution where the players can get the maximum sum of available payoffs
in this game. For the classical
correlation, this sum equals to 1.75, which is much smaller than
the value of 5 obtained with the quantum correlation.

\section{The effects of shared quantum and classical correlations on the dynamics of the sequential game}

The quantum version of MH is played as follows \cite{monty}: Three doors are
represented by three orthogonal bases of a qutrit,
$\ket{0},\ket{1},\ket{2}$, respectively. The system is described by
three qutrits as $\ket{\Psi}=\ket{o}_{O}\ket{b}_{B}\ket{a}_{A}$, where
$\ket{a}_{A}$ is Alice's choice of the door, $\ket{b}_{B}$ is
Bob's choice of the door, and $\ket{o}_{O}$ is the door that has been
opened. The sequence of the game is represented by
\beq\ket{\Psi_{final}} = (\hat{S}\cos{\gamma} +
\hat{N}\sin{\gamma})\hat{O}(\hat{I} \otimes \hat{B} \otimes
\hat{A})\ket{\Psi_{inital}}, 
\eeq 
where $\hat{A}$, $\hat{B} \in
SU(3)$ are Alice's and Bob's operators, respectively, $\hat{O}$ is
the opening operator, and $\hat{S},\hat{N}$ are the switching and the 
no-switching operators, respectively. Bob's wining probability is
$\$_{B}=\sum_{i,j} \abs{\braket{iij}{\Psi_{final}}}^2$.

Suppose that the initial state is a maximally entangled state between Alice and Bob,
$\ket{\Psi_{inital}}=\ket{0}_{O}\otimes
\frac{1}{\sqrt{3}}(\ket{0}_{B}\ket{0}_{A} + \ket{1}_{B}\ket{1}_{A}
+\ket{2}_{B}\ket{2}_{A})$. According to Ref. \cite{monty} , when Bob's
strategy is limited to a classical one, $\hat{\sigma_{0}}$, Bob's
payoff is given by

\begin{eqnarray}
\$_{B} & = & \frac{1}{3}{\sin}^2 \gamma (\abs{a_{00}}^2 + \abs{a_{11}}^2 + \abs{a_{22}}^2) \nonumber \\
       & + & \frac{1}{3} {\cos}^2 \gamma (\abs{a_{01}^2} + \abs{a_{02}^2} + \abs{a_{10}^2} + \abs{a_{12}^2} + \abs{a_{20}^2} + \abs{a_{21}^2} ),
\label{BsP:eqn}
\end{eqnarray}
where $\hat{A}=\{ a_{ij} \}$. Alice can  make her payoff
$\frac{1}{2}$ by choosing some specific strategy $\hat{H}$ in Ref. \cite{monty}. In fact, Alice
can make the game fair with the help of the entangled
state. On the other hand, if Alice's strategy is restricted to
$\hat{\sigma}_{0}$, Bob can always win the prize with the choice of no-switch.

In this study, we considered the case where the initial state is the
classically correlated state, $\rho_{initial}= \proj{0}\otimes\frac{1}{3}(\proj{00}
+ \proj{11} + \proj{22})$. We computed Bob's payoff with the
classically correlated state $\rho_{initial}$ and considered the
case where Bob's strategy is restricted to $\hat{\sigma}_{0}$.
Surprisingly, the same payoff as Eq.(\ref{BsP:eqn}) is obtained.
Consequently, Alice does not need the entangled state to make the game fair.

\section{Conclusion and Discussion}

We found that QO and QC can help players
resolve the dilemmas in the games except MP. In MP 
with QO and CC, the players cannot find an NE. However,
replacing QC with CC creates an NE
which gives zero payoff to players as is the case with classical
mixed strategy. It is worth investigating whether the fact that an NE disappears in the case of QC is intrinsic to zero sum games.

It is interesting to note that in all the symmetric games, 
the dilemmas are resolved with the same strategies,
$(i \hat{\sigma}_{z}, i \hat{\sigma}_{z})$. Another interesting
observation here is that the only games in which a unique NE is
achieved for QO and CC are SD and MP.
However, it must be noted that the payoffs obtained with CC are
much smaller than those obtained with QC in all the games except MP.
In the analyzed non-zero sum games, QC enables the players to
resolve their dilemmas with the highest possible $ \$_{A}+\$_{B}$ in
these games.

We have shown that the payoffs and the game dynamics are very much
affected by the types of correlations. This study revealed the basic
effects of correlations in game theory by comparing a specific
form of QC and one type of CC that is generated through phase
damping processes on the type of QC.
An important future direction is to examine to what extent more general
form of CC can simulate the results obtained with QC in game theory.

\section*{Acknowledgements}
The authors thank T. Yamamoto for stimulating discussions.

\end{document}